\documentclass[twocolumn,showpacs,preprintnumbers,amsmath,amssymb,superscriptaddress,prl]{revtex4}
\usepackage{dcolumn}
\usepackage{bm}
\usepackage[latin1]{inputenc}
\usepackage[dvips]{graphicx}
\usepackage{float}
\usepackage[dvips]{epsfig}
\usepackage[latin1]{inputenc}

\begin{document}

\title{Exciton/plasmon polaritons in GaAs/GaAlAs heterostructures near a metallic layer}

\author{J. Bellessa}
\email{bellessa@lpmcn.univ-lyon1.fr} \affiliation{Groupe d'Etude
des Semiconducteurs, Université Montpellier II, CNRS, Case
Courrier 074, F-34095 Montpellier Cedex 5, France}
\author{C. Symonds}
\author{C. Meynaud}
\author{J.C. Plenet}
\affiliation{Universit\'e de Lyon, Lyon, F-69003, France;
Laboratoire de Physique de la Mati\`ere condens\'ee et
Nanostructures, CNRS UMR 5586, Univ. Lyon-1, Villeurbanne F-69622,
France}
\author{E. Cambril}
\author{A. Miard}
\author{L. Ferlazzo}
\author{A. Lema\^{\i}tre}
\affiliation{Laboratoire de Photonique et de Nanostructures, CNRS,
Route de Nozay, F-91460 Marcoussis, France}

\date{\today}

\begin{abstract}

We report on the strong coupling between inorganic quantum well
excitons and surface plasmons. For that purpose a corrugated
silver film was deposited on the top of a heterostructure
consisting of GaAs/GaAlAs quantum wells. The formation of plasmon/
heavy-hole exciton/light-hole exciton mixed states is demonstrated
with reflectometry experiments. The interaction energies amount to
21 meV for the plasmon/light-hole exciton and 22 meV for the
plasmon/heavy-hole exciton. Some particularities of the
plasmon-exciton coupling were also discussed and qualitatively
related to the plasmon polarization.

\end{abstract}

\maketitle Since its first demonstration by Weisbuch et al.
\cite{weisbuch}, the strong photon - exciton coupling regime has
been extensively studied in inorganic semiconductor planar
microcavities \cite{skolnick}. In this regime the excitons
hybridize with the microcavity photons to form polaritons.
Epitaxial inorganic semiconductors are materials of choice for
these studies since their well-mastered growth yields layers of
high crystalline quality. In particular, quantum wells (QWs)
exhibit intense excitonic transitions with narrow linewidth
suitable for reaching the strong coupling regime. More recently,
polariton non linearities \cite{savvidis} and Bose-Einstein
condensation have been claimed in semiconductor QWs coupled to
planar microcavities \cite{kasprzak}, as well as  polariton lasing
in GaN microcavities at room temperature \cite{christopoulos} and
in micropillar GaAs/GaAlAs cavities~\cite{Aristide}.

The strong coupling regime has been also demonstrated with surface
plasmons (SPs) as predominant optical modes \cite{bellessa}. This
regime has been obtained in organic semiconductors, with very
large Rabi splitting (several hundreds of meV). During the past
decade, the improvement of metal nanostructuration has lead to a
renewed interest in SPs \cite{barnes}. For example, SPs photonics
bandgaps \cite{Kitson}, low losses SPs guides and interferometric
systems \cite{BozhevolnyiPRL,BozhevolnyiNat} have been
successfully developed. The interactions between SP and emitters
have been also intensively studied. In the weak coupling regime, a
two orders of magnitude enhancement of the spontaneous emission
rate was reported, leading to improved emitting devices
\cite{neogi,okamoto}, and efficient coupling between single
quantum boxes and metallic wires have been demonstrated
\cite{Akimov}. In the strong coupling regime, the metal
structuration has been used to tailor the properties of the
plasmon/exciton mixed states \cite{sugawara, wurtz, Dintinger} and
for efficient polariton extraction \cite{BonnandCC}. Until now, in
the strong coupling regime, the experiments were restricted to
organic materials, J-aggregated dyes, or hybrid organic/inorganic
materials (perovskites) \cite{symonds}. These materials have high
oscillator strength and can be used at room temperature. They are
however extremely sensitive to the environment, easily
deteriorated by light irradiation, and their structural properties
are less controlled than those of inorganic semiconductors.

In this paper we show that the strong coupling regime can be
achieved between SPs and GaAs/AlAs QW excitons. The wide
possibilities of plasmon mode engineering coupled to the highly
controlled epitaxial heterostructures make this type of structures
promising for future developments. In the first part of the paper,
the sample and the experiment are described. The dispersion
relations deduced from the reflectometry experiments are analyzed
in a second part. The formation of coupled plasmon/exciton mixed
states is demonstrated. In particular, we show that these
polariton states are an admixture of both heavy- and light-hole
excitons and SPs. Finally typical behaviors of plasmon/exciton
polaritons are qualitatively discussed, considering the
polarization of the plasmon modes in interaction with anisotropic
structures.

The sample is formed by a corrugated silver film deposited on a
GaAs/GaAlAs QW structure. The heterostructure was grown by
molecular beam epitaxy on a GaAs [001] substrate. It consists of
five GaAs QWs (10 nm thick) separated by Al$_{0.93}$Ga$_{0.07}$As
barriers (15 nm). The last QW was covered by a 27 nm thick
Al$_{0.93}$Ga$_{0.07}$As barrier and a 3 nm GaAs cap. The upper
barrier is larger to prevent non radiative recombinations induced
by surface states or by the semiconductor-metal interface
\cite{marzin}. A luminescence spectrum of the heterostructure
without the silver layer is shown in Fig. 1. The QWs were excited
with the 488 nm line of an argon laser and maintained at a
temperature of 77~K. Two peaks are observed at 1554 and 1570 meV
with a full width at half maximum of 3 and 4 meV, respectively.
They correspond respectively to the heavy-hole ($X_{hh}$) and the
light-hole ($X_{lh}$) excitons.

\begin{figure}
\includegraphics[width=7.0cm]{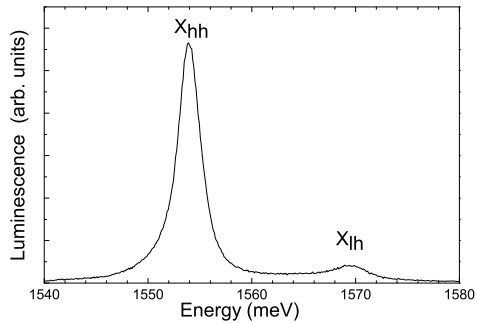}
\caption{\label{fig:epsart} Luminescence spectrum of the
semiconductor heterostructure without silver film, measured at 77~K.
}
\end{figure}

\begin{figure}
\includegraphics[width=8.5cm]{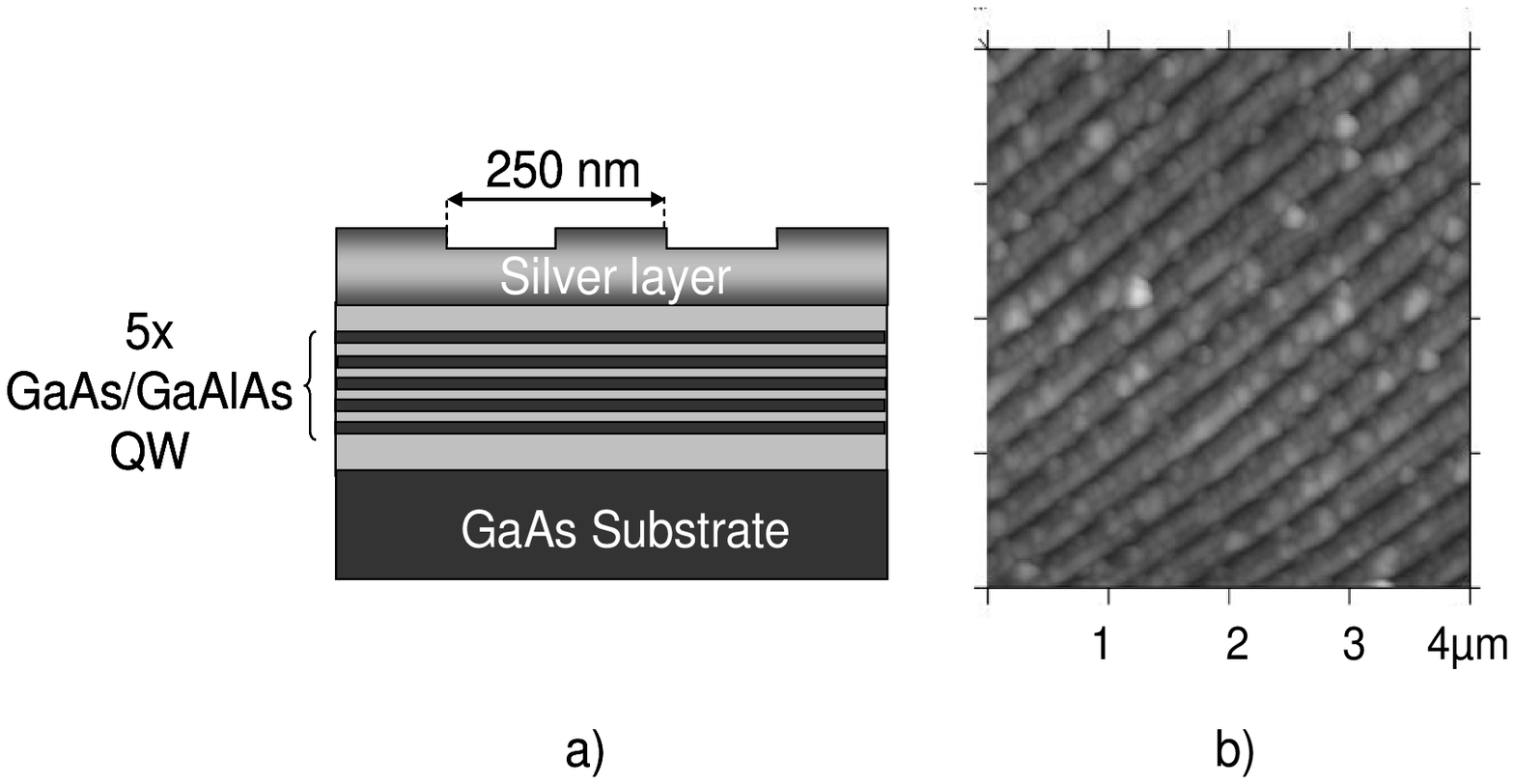}
\caption{\label{fig:epsart}(a) Schematic layout of the sample
showing the heterostructure grown on a GaAs substrate and caped by
an engraved silver film (not to scale). (b) Image of the surface
topography of the engraved silver film recorded by Atomic Force
Microscopy.}
\end{figure}

Two distinct SP modes coexist on both interfaces of the silver
film: Ag/semiconductor and Ag/vacuum. For a flat metal surface,
these modes cannot couple to the free-space radiative modes.
Therefore a grating has been engraved on the silver surface to
make this coupling possible and hence to give access to the
dispersion relations. The period of the grating was chosen so that
the Ag/semiconductor plasmon couples to free space radiations via
first order scattering. For the Ag/vacuum plasmon, no diffracted
order couples to free space modes in the energy range investigated
in this study. This geometry prevents the apparition of Ag/vacuum
plasmon peaks in the optical spectra and possible interactions
between Ag/semiconductor and Ag/vacuum plasmons \cite{barnes2}.

The fabrication of the engraved silver film was done as follows:
first a thin Ag film (60 nm thick measured with a quartz balance
during the evaporation) was deposited on the sample. The grating
was then defined by e-beam lithography and partially engraved by
Ar ion beam etching. A layout of the sample is drawn in the Fig.
2(a). The surface of the engraved silver film was imaged with an
atomic force microscope and is shown in Fig. 2(b). The measured
grating period is $250\pm$10~nm and the groove depth is 20 nm.

Angular resolved reflectometry experiments have been performed
with the sample mounted on the cold finger of a nitrogen cryostat,
cooled at 77~K. A white light source is focused on the sample. The
reflected light is detected by a spectrometer coupled to a cooled
Si detector with a lock-in amplifier. A goniometer is used to tune
the reflection angle. Reflectometry spectra recorded at different
angles are shown in the Fig. 3. At $\theta=29^{\circ}$, three well
separated dips are present in the spectrum. The large dip at 1482
meV is related to the SP mode. The two sharp dips, well separated
from the SP line and lying at 1555 and 1573~meV, correspond to the
heavy- and light-hole excitons. The Stokes shift is 1~meV (resp.
3~meV) for the heavy- (resp. light-) hole excitons, which is
typical for good quality QWs.

\begin{figure}
\includegraphics[width=7.0cm]{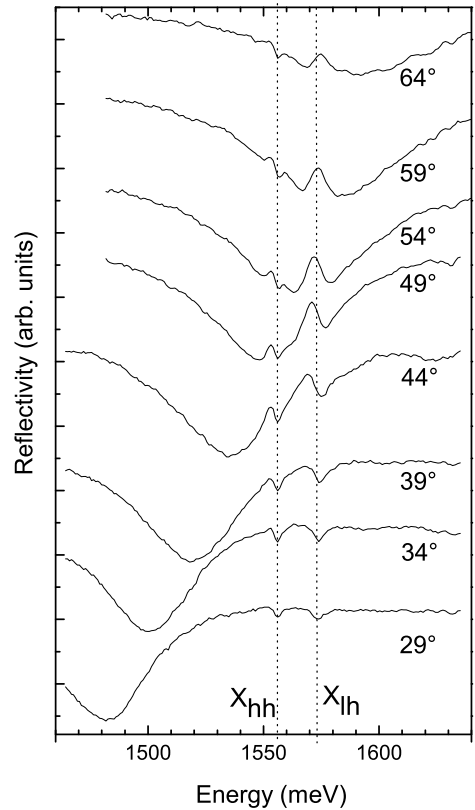}
\caption{\label{fig:epsart} Reflectometry spectra as a function of
the incident light energy for different detection angles. The
spectra are arbitrary translated for the clarity of the figure.
The dashed lines at 1555 and 1573 meV indicates the energies
corresponding to the heavy- and light-hole excitonic transitions
respectively.}
\end{figure}

Several spectral modifications occur as the detection angle
increases. At small angles ($\theta\leq39^{\circ}$) the SP energy
increases because of the SP dispersion, while the excitons
energies remain almost constant. Above $44^{\circ }$, when the SP
energy is close to the QW excitons energies, the peak previously
associated to the light-hole exciton at small angles (1573 meV)
starts to move toward higher energies and progressively broadens.
The evolution of the dip associated to the heavy-hole exciton is
more complex: while a dip remains visible at 1555 meV at all
detection angles, a second dip is observed remaining on the low
energy side of the 1573 meV dip up to $59^{\circ}$. A better
understanding is gained when plotting the dip energies as a
function of the in plane wavevector of the detected light
($k_{d}=\frac{2\pi}{\lambda}\sin(\theta)$), as in Fig. 4.

As mentioned above, a dip is always present at 1555~meV. On the
opposite, the energy of the dip associated to the SP mode at small
wavevector first increases, as expected for a bare SP mode, but
then saturates at a value slightly below 1555~meV for large
wavevector; whereas the energy of the dip at 1573~meV remains
constant at $k\le5.5~\mu m^{-1}$ and increases at higher
wavevector to follow the bare SP dispersion. Last, a fourth dip
appears around $k=6~\mu m^{-1}$, emerging from the 1555 meV dip
and moving asymptotically toward the bare light-hole exciton
energy. These behaviors are the signature of two successive
anticrossing, unveiling the strong coupling between the plasmon
mode and both excitons.

\begin{figure}
\includegraphics[width=8.5cm]{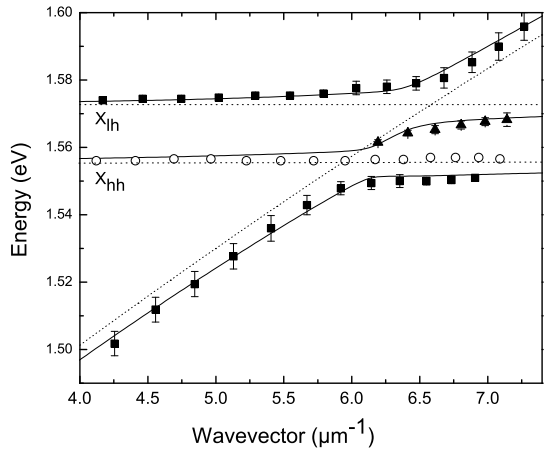}
\caption{\label{fig:epsart} Reflectometry dips energy
($\blacksquare$, $\blacktriangle$ and $\circ $) as a function of
the in-plane wavevector of the detected light. The dashed line is
the dispersion relation of a bare SP calculated with a transfer
matrix method. The doted lines are the X$_{hh}$, X$_{lh}$ bare
energies. In full lines are drawn the calculated polaritons
energies.}
\end{figure}

In order to analyze the experimental results, the
coupled $X_{hh}$/$X_{lh}$/plasmon polariton dispersions were
calculated using a coupled oscillator model. The wavevector in the
layer plane is a good quantum number. The dispersions were obtained
by diagonalizing the exciton-plasmon Hamiltonian \cite{sermage}:

\begin{center}
$H=\left(
   \begin{array}{ccc}
     E_{pl}(k)-i \gamma_{pl} & V_{Xhh}/2 & V_{Xlh}/2 \\
     V_{Xhh}/2  & E_{Xhh}-i \gamma_{Xhh} & 0 \\
     V_{Xlh}/2  & 0 & E_{Xlh}-i \gamma_{Xlh} \\
   \end{array}
 \right)
$
\end{center}

where $E_{pl}(k)$ is the bare surface plasmon energy,
$\gamma_{pl}$ the plasmon homogeneous broadening, $E_{Xhh}$ and
$E_{Xlh}$ the bare heavy- and light-hole exciton energies,
$\gamma_{Xhh}$ and $\gamma_{Xlh}$ the heavy- and light-hole
exciton linewidth (3 and 4 meV, respectively) and $V_{Xhh}$
($V_{Xlh}$) the interaction energy between the plasmon and the
heavy-hole exciton (light-hole exciton). In the considered
wavevector range, both excitons energies can be regarded as
dispersionless. The energy $E_{\text{plasmon}}(k)$ is calculated
using a conventional transfer matrix method for the semiconductor
structure described above with a 50~nm flat silver layer (mean
thickness value of the corrugated layer). The plasmon broadening
deduced from the calculations is 25~meV and is taken as
homogeneous width of the plasmon. The inhomogeneous broadening
comes from variations of the grating period and of the metal
interface in the large sample surface studied in reflectometry
experiments ($5~mm^{2}$). The wavevector $k_{d}$ of the SP first
diffracted order is given by:
\begin{center}
$k_{\text{d}}=k-\frac{2\pi}{a}$
\end{center}
where $a$ is the silver film grating period. A period of 245 nm
was used. The bare plasmon dispersion curve fits well the
experimental data far from the resonance where the interaction
with the exciton is negligible. Three $k$-dependant polariton
energies are obtained. The grating first diffracted order of the
calculated polaritons dispersion lines are shown in the Fig. 4
(energies as a function of $k-2\pi/a$). The interaction energies
$V_{Xhh}=22\pm 2$ meV and $V_{Xlh}=21\pm 2$ meV have been obtained
by fitting the reflectometry dips energies, and the calculated
curves are in good agreement with the experimental curves. This
demonstrates that the reflectometry dips result from the mixing of
plasmon and excitons and that a strong coupling regime can be
obtained for inorganic quantum wells in the vicinity of a metal
film. The interaction energies are of the same order of magnitude
than in semiconductor microcavities in strong coupling
regime~\cite{skolnick}.

The wavefunction of the polaritons formed is an admixture of bare
plasmon, heavy- and light-hole excitons with $|\phi\rangle =
\alpha_{p} |\text{plasmon}\rangle +\alpha_{hh}|X_{hh}\rangle
+\alpha_{lh}|X_{lh}\rangle$. In order to quantify this mixing, the
fraction of bare modes of the middle polaritons line are shown on
the Fig. 5. A large mixing between $X_{hh}$ and $X_{lh}$ excitons
is induced by the coupling to the SPs. A state with comparable
weight of plasmon, heavy- and light-hole excitons is obtained at
$k=6.3 \mu m^{-1}$  ($|\alpha_{p}|^2=0.36, |\alpha_{lh}|^2=
|\alpha_{lh}|^2=0.32 $).

\begin{figure}
\includegraphics[width=8cm]{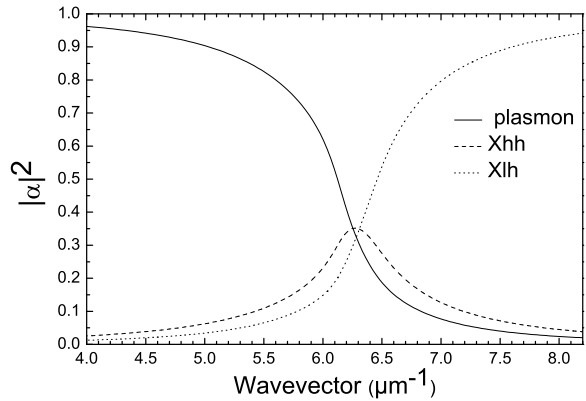}
\caption{\label{fig:epsart} Middle branch polariton wavevector
coefficient for the bare plasmon and excitons, deduced from the
coupled oscillator model.}
\end{figure}

The three polaritonic energies are well reproduced by the coupled
oscillator model, but not the quasi dispersionless line at the
bare heavy-hole exciton energy (open circles on the Fig. 4). The
coexistence of the signature of polaritons and incoherent states
(states not coherently coupled to plasmons \cite{agranovitch}) has
already been observed for organic semiconductors near metal layers
in luminescence experiments \cite{bellessa}. In semiconductor
microcavities, in the strong coupling regime, the incoherent
luminescence may appear at the bare exciton energy with low
reflectivity mirrors \cite{Hobson} but not with high reflectivity
Bragg mirrors \cite{houdre}. In our case, the dispersionless line
can be related to transparency modes through the silver film since
a transmission of a few percents is expected.

The signature of the absorption of the heavy-hole exciton and
light-hole exciton can be seen by transparency at an angle of 29°,
when the plasmon is well separated from the exciton. The
plasmon/exciton polaritons and the transparency modes are detected
in the same direction but do not have the same wavevector. The
transparency mode has the same in plane wavevector as the detected
light whereas the polariton is first order diffracted. When the
detection angle increases, the behavior for the heavy- and
light-hole exciton differs : the incoherent line remains present
during the anticrossing for the heavy-hole exciton but disappears
for the light-hole exciton. This particular feature of plasmon in
strong coupling can come from the polarization of the SP mode.
Indeed, this surface wave only exists in TM polarization and the
electric field is mainly transverse (perpendicular to the layer
plane) \cite{polar}. The transparency modes propagate in a
direction quasi perpendicular to the layer plane (10° in the
semiconductor for a detection angle of 40°) and the corresponding
electric field is in this plane. Since optical transitions are
forbidden for heavy-hole excitons when the field is perpendicular
to the layer plane \cite{bastard}, the interaction is reduced with
the plasmon (only the longitudinal electric field interacts) but
not with the transparency mode. Therefore the dip related to the
incoherent states is still present during the anticrossing. For
the light-hole exciton, on the opposite, the transition is allowed
for the electric fields both perpendicular and parallel to the
layer plane, i.e. for both longitudinal and transverse components
of surface plasmon and for the incoherent mode. In this case the
incoherent dip is masked when the plasmon anticrosses the exciton.

Finally it has to be noticed that an adequate grating is necessary to
observe the plasmon/exciton polariton. When a semiconductor
heterostructure is close to a flat metal surface, the polariton
can not couple to free space radiative modes. The
quantum well excitons can be strongly coupled to SP, inducing coupling
between the different excitons and dynamic modification, but would not
be detectable directly without modification of the upper metal interface.

In conclusion we have shown that the strong coupling between a
surface plasmon and the excitons of GaAs/GaAlAs quantum wells
occurs when the quantum wells are located in the vicinity of the
metal surface. Polaritons are formed by the mixing of plasmon,
heavy- and light-hole exciton states. The interaction energies are
22~meV and 21~meV for the heavy- and light-hole exciton
respectively. Finally we observed some specific behaviors in the
exciton-plasmon coupling related to the polarization of the
electric field associated with surface plasmons.

This work has been supported by the project SCOP from the french
Agence Nationale de la Recherche.


\begin{thebibliography}{}
\bibitem{polar} In the energy range considered, an order of magnitude of the ratio of the amplitude of the longitudinal electric field on the transverse component is 0.6.
\end{thebibliography}
\end{document}